# An Empirical Study on the Effectiveness of Data Resampling Approaches for Cross-Project Software Defect Prediction

Kwabena Ebo Bennin[1], Amjed Tahir[2], Stephen G. MacDonell[3], Jürgen Börstler[4(✉)]

[1]*Information Technology, Wageningen University and Research, Wageningen, the Netherlands*
[2]*Massey University, Palmerston North, New Zealand*
[3]*University of Otago and Auckland University of Technology, Dunedin, New Zealand*
[4]*Blekinge Institute of Technology, Karlskrona, Sweden. email: jurgen.borstler@bth.se*

**Abstract**

*Cross-project defect prediction (CPDP), where data from different software projects are used to predict defects, has been proposed as a way to provide data for software projects that lack historical data. Evaluations of CPDP models using the Nearest Neighbour (NN) Filter approach have shown promising results in recent studies. A key challenge with defect-prediction datasets is class imbalance, that is, highly skewed datasets where non-buggy modules dominate the buggy modules. In the past, data resampling approaches have been applied to within-projects defect prediction models to help alleviate the negative effects of class imbalance in the datasets. To address the class imbalance issue in CPDP, the authors assess the impact of data resampling approaches on CPDP models after the NN Filter is applied. The impact on prediction performance of five oversampling approaches (MAHAKIL, SMOTE, Borderline-SMOTE, Random Oversampling and ADASYN) and three undersampling approaches (Random Undersampling, Tomek Links and One-sided selection) is investigated and results are compared to approaches without data resampling. The authors examined six defect prediction models on 34 datasets extracted from the PROMISE repository. The authors' results show that there is a significant positive effect of data resampling on CPDP performance, suggesting that software quality teams and researchers should consider applying data resampling approaches for improved recall (pd) and g-measure prediction performance. However, if the goal is to improve precision and reduce false alarm (pf) then data resampling approaches should be avoided.*

## 1. INTRODUCTION

Defect prediction models can help to identify defective software components and thereby support managers in resource allocation. Previous studies have shown that defect prediction models can yield useful results [1, 2], but their reliability might be affected by the quality of the underlying datasets [3] or confounding variables affecting the measures used for creating the prediction models [4, 5, 6]. Several studies have proposed prediction models based on different statistical and machine-learning approaches [7, 8, 9]. However, the performance on these models largely depends on historical data, obtained either from the same project (in case historical data exist) or from projects that are very similar to the project under consideration regarding content and context.

Zimmermann et al. [10] cautioned that defect prediction models perform well within projects as long as enough data for training of the prediction models exist. However, for new and unfamiliar projects, the lack of historical data becomes a challenge. A promising approach to handle this issue is to use a cross-company or cross-project defect prediction (CPDP), where data from other companies or projects are used for model training. To help in obtaining the most suitable training data for CPDP, different techniques have been proposed and validated, including data filtering techniques such as Nearest Neighbour (NN) filter [11], Double Transfer Boosting [12], and clustering [13, 14].

NN filter, a data filtering approach, has been shown to perform significantly better than several ensemble, boosting or transfer-learning-based approaches [15]. Hosseini et al. [16] and Turhan et al. [17] confirmed that the NN filter can have a positive impact on the performance of CPDP models. NN filter eliminates irrelevant data instances based on the characteristics of the target distribution, selecting only the more suitable defective and clean instances [11].

However, class imbalance is a prevalent problem in data mining and defect prediction [18, 19], where the majority of the instances are clean or not faulty [19, 20]. Consequently, the acquired dataset for cross-project model training will most likely be highly skewed towards one class (that is the non-defective or clean instance). Resampling approaches such as simple Random Over-Sampling (ROS), Random Under-Sampling (RUS), and synthetic methods such as SMOTE have been proposed to alleviate the negative effects of class imbalance on the performance of Within Project Defect Prediction (WPDP) models [21]. Resampling approaches aim to increase the number of minority class samples (defective modules), and they can significantly improve the performance (recall, g-mean) of defect prediction models [22].

The potential benefit of data resampling approaches in mitigating the negative effect of class imbalance on CPDP models has been investigated by few researchers. Previous studies [23, 12, 24, 25] considered the effect of very few data resampling approaches or evaluate the performances of data resampling approaches on CPDP using a few datasets. Other studies propose complex methods integrated with data resampling approaches to improve CPDP performances [26, 27]. A recent systematic literature review by Hosseini et al. [16] revealed that most studies in CPDP fail to use multiple performance measures and fail to apply robust statistical tests, including effect sizes, which resulted in an unfair comparison of the performances of CPDP to WPDP models.

The motivation of this study is to augment the few existing studies and examine the practical benefits of data resampling approaches on CPDP models. We assess the impact of applying eight commonly used data resampling methods (MAHAKIL, SMOTE, Borderline-SMOTE, ADASYN, ROS, RUS, Tomek links and OSS) in the domain of CPDP after acquiring data from different projects (using the NN filter) and use the resampled data datasets for the training of the defect prediction model for a different project. Additionally, we conduct robust statistical tests on the results by testing for statistical significance using Brunner's statistical test [28] and apply Cliff's effect size to examine the practical benefits of the applied data resampling approaches. We selected the NN filter because it is easy to implement compared to other CPDP approaches and has been shown to improve CPDP performance in previous studies [29, 14, 16, 15]. To assess the impact of data resampling approaches on NN-filtered datasets, this study explores the following research questions.

1. RQ1: What is the impact of data resampling approaches on NN-filtered datasets in CPDP?
2. RQ2: What are the high-performing resampling approaches for NN-filtered datasets in CPDP?
3. RQ3: Is the application of data resampling approaches practical for CPDP?

The contributions of this paper are as follows:

- A benchmark experiment that shows that recall ( $pd$ ) and *g-mean* performances of CPDP can be improved by applying data resampling approaches to NN-filtered datasets.
- A demonstration that *oversampling* and *random undersampling* methods always produce higher false alarms ($pf$) in CPDP. NN-filtered data with no resampling produces the best pf values.
- A python package for MAHAKIL—an easy-to-use tool for oversampling class imbalanced data.

The remainder of this paper is structured as follows: Section 2 presents the related work. In Section 3, we discuss our methods and experimental settings. Our results and a discussion of the results are reported in Sections 5 and 6, followed by a discussion of potential threats to validity in Section 7. Finally, we present our conclusion from this study with potential future work directions in Section 8.

## 2. RELATED WORK
### 2.1 Cross-project defect prediction

One of the first to attempt at developing CPDP models was Zimmermann et al. [10]. By conducting a large-scale experiment on 12 real-world datasets, 622 cross-project prediction models were analysed and investigated for the feasibility of cross-company defect prediction (CCDP) models. After observing a low success rate of 3.4%, they conclude that CCDP is still a challenge. Turhan et al. [11] proposed a practical defect prediction approach for organizations aiming to employ defect prediction but that lacks historical data. Applying the principles of analogy-based learning, they use the *k-nearest* neighbour algorithm to select 10 nearest data instances from other company data for every unlabelled test instance for CCDP. They demonstrated that even small data samples acquired using their approach could be used to build effective defect predictors. He et al. [30] conducted a large study using open-source projects to investigate the feasibility of CPDP, considering careful data selection approaches. The obtained results were similar to those achieved by previous studies [10, 11], indicating that CPDP works well in a few cases and carefully selecting the training data improves prediction performance though not necessarily selecting projects in the same domain. A recent benchmark study by Herbold et al. [15] complemented the positive impact of the data filtering approaches on CPDP models.

Other studies, including [31, 12, 32, 33], have employed several transfer learning and data boosting techniques to improve CPDP performance. Considering two projects, Watanabe et al. [34] conducted an inter-project prediction and demonstrated that data characteristics had an impact on the success of cross-project predictions. They adapted CPDP by using a metric compensation method that adjusted the average values of each metric in both the training and test set to the same level and achieved high precision and recall values. Transfer learning techniques have also been applied in the domain of CPDP. In order to make the feature distribution of the source projects and target projects similar, Nam et al. [31] applied Transfer Component Analysis, which transforms the data based on the new feature representation discovered from both the source and target projects. They found that the prediction performance increased significantly after experimenting on eight open-source projects. Similarly, Yu et al. [32] proposed a novel semi-supervised clustering-based data filtering method that filters the data and adopts multi-source TrAdaBoost algorithm, an effective transfer learning method, into cross-company prediction to import knowledge not from one but from multiple sources to avoid a negative transfer. Poon et al. [35] proposed a Credibility theory-based Naive Bayes (CNB) classifier that uses a novel re-weighting mechanism to adapt the source data to the target data distribution simultaneously. The method ensures that the pattern of the source data and experimental results improved the performance over other CPDP methods. The study of Zhou et al. [36] investigated the performances of a number of CPDP techniques and models with simple size models and observed that simple size models in most cases outperformed the complex and recently proposed CPDP techniques. Asano et al. [37] applied bandit algorithms to help in selecting the most suitable projects for CPDP models. Our study aims to complement prior studies by aiming to improve the performance of existing CPDP models that adopts the NN filter approach.

**2.2 Data resampling application in CPDP**
Although the class imbalance issue is known to be critical for defect prediction models, only a few studies have discussed the challenge of class imbalance in the context of CPDP. Ryu et al. [26, 27] proposed two methods, a boosting and instance weighting technique that uses transfer learning to solve the class imbalance issue. Jing et al. [38] employed a semi- supervised transfer component analysis to balance the source and target datasets before applying their proposed semi- supervised transfer component analysis and improved subclass discriminant analysis for prediction.

Most of the previous studies' aim was to improve prediction performance by integrating data resampling approaches as part of their solutions. A study by Kamei et al. [25] on Just-In-Time (JIT) defect prediction models using 11 open-source cross-projects revealed that JIT models rarely improved in general, but did improve when selected training projects are similar to the testing data, larger set of training data is provided or using ensemble models. The random undersampling approach was applied to the training datasets before training the JIT models. Other studies have adopted the popular SMOTE approach in alleviating the negative effects of class imbalance. Chen et al. [12] used the SMOTE resampling approach to balance their training datasets before applying their proposed Double Transfer Boosting model. Recently, Limsettho et al. [29] proposed CDE-SMOTE that adopts the SMOTE data resampling approach to alleviate the negative effects of class distribution mismatch and imbalance between the training and test datasets. The authors estimate the proportion of each class of the target dataset using class distribution estimation and employ the SMOTE method to generate data samples equal to the estimated amount. Additionally, Goel et al. [23] observed that SMOTE improves the performance of CPDP models. Yu et al. [24] were the first to conduct an empirical study comprising of more than one data resampling approach. They applied six data resampling approaches on 15 open-source datasets and three CPDP models and observed that the performance improved when data resampling approaches are applied. More information on CPDP studies can be found in the recent literature review and meta-analysis study conducted by Hosseini et al. [16]. The comprehensive studies of Bennin et al. [21] and Tantitham-thavorn et al. [39] provide more information on the impact of data resampling approaches on software defect prediction models.

Previous studies aim to improve CPDP performance by using complex boosting and transfer learning techniques. Most studies [23, 12, 29, 25] investigated the effects of a single type of data resampling approach on CPDP performances. The study presented in this paper differentiates itself from the studies discussed above, and specifically, the study of Yu et al, [24], in which it applies eight data resampling approaches to the acquired data after applying an NN filter and empirically investigate the impact of data resampling approaches on the performance of CPDP. Compared to WPDP, filter-based CPDP approaches including the NN filter try to generate a project dataset that is similar to a target project dataset. This NN filter procedure is a simple approach that is independent of the prediction model, straightforward to implement and has been shown to have a positive impact on CPDP models [17, 15, 16].

# 3. BACKGROUND
In this section, we present a brief description of the NN filter and an overview of the data resampling approaches we employed in our experiments. The NN filter is selected because (1) it is easier to implement, compared to other similar approaches, (2) it has been widely (and successfully) used in previous studies [29, 14, 40] and (3) it is known to improve prediction performance in comparison to other data filtering techniques [16, 15].

**3.1 NN filter**
This approach proposed by Turhan et al. [11] is a relevancy filter that effectively selects the closest data instances with respect to the new target project from a collection of various projects based on the K-NN algorithm. It is a pre-processing approach that could be combined with the normal classification process. It was combined with a Naive Bayes classifier by the same authors. The procedure for NN filtering is as follows:

1. Aggregate and merge all training data into one set of data.
2. For each module in the target project, find k neighbours from the combined training data considering their pairwise
3. Euclidean distances.
4. Collect the selected neighbours and remove duplicated data instances to obtain the filtered dataset.

**3.2 Oversampling approaches**
*3.2.1 Random oversampling (ROS)*
ROS can be considered as one of the most basic oversampling approaches. Minority instances are randomly selected and replicated, and thus, no new information is provided for the classifier. ROS is very easy to implement and has been widely used in several studies in defect prediction [41, 19, 21, 42].

*3.2.2 Synthetic Minority Oversampling Technique (SMOTE)*
A technique designed to alleviate the effect of class imbalance on prediction performance [43]. Several defect prediction studies [44, 45, 46] have adopted and applied SMOTE in the literature. SMOTE generates 'synthetic' minority samples by considering each minority class sample. Based on the parameter values ($k$), new synthetic samples are created along with the line segments that join $k$ minority class NNs of each sample under consideration.

*3.2.3 MAHAKIL*
A recently developed state-of-the-art oversampling technique that aims to generate diverse minority class samples in order to improve prediction performance and reduce false alarms [22]. Different from the conventional $k$ NN methods, MAHAKIL utilizes the chromosomal theory of inheritance to generate diverse synthetic data instances. MAHAKIL alleviates the overgeneralization of prediction models since synthetic data instances created are not clustered into a specific region of the minority class of the dataset. The code used in this study (a python package) is provided online for future use and replication studies[1]

*3.2.4 Adaptive Synthetic Sampling (ADASYN)*
By focusing only on the minority class samples that are difficult to classify, the ADASYN approach [47] assigns weights to the minority classes and dynamically adjusts the weights in a bid to reduce the bias in the imbalanced dataset by considering the characteristics of the data distribution. The ADASYN algorithm incorporates a density distribution in automatically deciding the number of synthetic samples needed for each minority class sample. In contrast to the SMOTE algorithm, which generates equal synthetic samples for each minority class, the ADASYN learning algorithm is induced to focus on the hard to learn examples within the minority class samples; therefore, samples generated are not equal for all samples.

*3.2.5 Borderline-SMOTE*
This technique is a modification of the original SMOTE [43] that focuses more on the instances that are harder-to-classify, that is, instances on the borderline of the classifier [48]. Instances that are marked as hard-to-classify are labelled and considered for generating new synthetic data using SMOTE. Borderline-SMOTE aims to clearly set the decision boundary for the trained classifier for improved prediction performance.

**3.3 Undersampling approaches**
*3.3.1 Random undersampling (RUS)*
RUS aims to balance the dataset by randomly selecting and deleting the majority of instances. RUS is easy and fast to implement. As a conventional technique, it is a widely used undersampling technique in several empirical studies [41, 21, 19, 42].

*3.3.2 Tomek links*
Tomek [49] observed that instances of different classes could be very close neighbours and overlap each other. These instances are closer to each other than they are to their own class samples, and they form what is referred to as Tomek links. The majority of instances that are part of the Tomek links are considered as noise and are deleted by this technique until all minimally distanced nearest-neighbour pairs are of the same class.

*3.3.3 One-sided selection*
Kubat et al. [50] proposed the undersampling approach called One-sided selection. This technique aims to remove noisy and borderline majority instances by adopting the concept of Tomek links [49]. The technique works by randomly selecting a majority sample and combining it with all minority instances to create a new set S. The new set S is used to classify the original dataset and all misclassified instances are added to S. It considers majority instances to be redundant if these instances in S participate in Tomek links.

# 4. EVALUATION AND EMPIRICAL ANALYSIS

The research questions and experimental design and execution are described in this section.

**4.1 Research questions and experiment design**
The goal of this study is to empirically examine and assess the impact of using data resampling approaches on NN-filtered datasets used in CPDP models. To achieve our goal, we carried out two sets of experiments. The experiments are conducted and executed to answer the following three research questions.

1. RQ1: What is the impact of data resampling approaches on NN-filtered datasets in CPDP?
2. RQ2: What are the high-performing resampling approaches for NN-filtered datasets in CPDP?
3. RQ3: Is the application of data resampling approaches practical for CPDP?

The research questions are used to examine the impact of applying several data resampling approaches on the performance of the CPDP. The study considered only defect- proneness (predicting if a module is defective or not) and not defect counts. We follow the criteria used by Peters et al. [51] and Zimmermann et al. [10] to determine if resampling approaches are beneficial and practical for CPDP. The work of Zimmermann et al. [10] demonstrated that a strong predictor is the one that achieves g-mean, recall and accuracy above 75%. By using a simple criterion, the total count of test-sets with predictors (models) that produced g-mean, recall and auc above 75% are computed. Since we include pf measures, we set it as practical if pf is less than 15%. Detailed description on the dataset used, experimental setup and analysis is discussed below.

**4.2 Datasets**
We obtained data from thirty-four (34) versions of twenty-two (22) open-source projects that are extracted from the PROMISE data repository, which are freely available[2]. Dataset size, imbalance ratio and projects with more than one version all carry heavyweight in our selection criteria, though we keep to datasets widely used in previous studies. Table 1 shows a summary of the description of these datasets. These projects, donated by Jureczko and Madeyski [13] and Jureczko and Spinellis [52], were written in Java where a module corresponds to a Java file. We only consider top-level classes (where the class has the same name as the source file, i.e. inner classes are ignored). The datasets vary in size and imbalance ratios, and they provide an extensive domain for evaluating the impact of resampling approaches and have been used in several previous studies. Twenty-one (21) code metrics from the CK and other metrics suites (see Table 2 for details) are collected using the BugInfo and ckjm[3] tools from each module. A non-defective module is labelled as zero, and defective modules are labelled with the number of bugs present in the module. Projects that had more than one version available were all merged to create our training (source) datasets. Single-version datasets (the first 14 projects) were thus used as the test (target) dataset where we applied the NN filter to selectively produce the filtered training datasets for each of the test datasets.

---
[1] https://bit.ly/3oVDji7

[2] https://zenodo.org/communities/seacraft

### 4.3 Evaluation measures

To assess our modules, we use four performance evaluation metrics: recall (i.e. probability of detection [$pd$]), *Area Under the ROC Curve (AUC)*, g-measure and *Probability of false alarm* ($pf$). These performance metrics range between the values of 0%–100%. We decided not to include precision and *F- measure* as they have been refuted as unstable for assessing imbalanced datasets [53]. Recall ($pd$) measures how much of the defective modules were detected.

In the general sense, a higher recall denotes better performance. The probability of false alarm ($pf$) measures the rate of wrongly predicted modules that were non-defective. A low or zero value for pf implies a better prediction model. For a highly imbalanced dataset, the AUC and g-measure (g-mean) performance metrics are preferred as they consider both the pf and pd values and do not value one metric over the other. AUC has also been recommended to be very stable for imbalance learning [21, 39]. For practical results, Zimmermann et al. [10] recommend pd values of 75% and above, which is maybe very challenging to achieve in CPDP studies. Herbold et al. [15] also observed that CPDP models rarely achieved a high recall of 75%. We use multiple performance measures to fully grasp the capabilities and dynamics of the examined prediction models and resampling methods as noted by the authors in [16]. The mathematical definitions of $pd$, $pf$ and $g\text{-}measure$ metrics are presented below. Computation of the AUC of ROC can be found elsewhere [54].

$$Recall(pd) = \frac{TP}{TP + FN} \quad (1)$$

$$pf = \frac{FP}{FP + TN} \quad (2)$$

$$g - measure = \frac{2 * pd * (100 - pf)}{pd + (100 - pf)} \quad (3)$$

### 4.4 Experimental setup

Two sets of empirical experiments are conducted. We first evaluate the performance of using an NN filter for CPDP, and we then investigate the influence data resampling approaches on the performance of CPDP models after filtering the training datasets. Implementation of the data resampling methods and the NN filter was executed using Imbalanced- learn [55], an open-source processing toolbox and the model construction and evaluation were executed using the open- source scikit learn [56] library available in python. We use all metrics from the datasets as suggested in [2]. For this study, five methods widely used in defect prediction studies were chosen. We considered Random Forests, Naive Bayes, K-NN algorithm, NNET and SVM [1]. Furthermore, we adopted a recently proposed boosting model called XGBoost [57]. XGBoost is an improved implementation of a gradient boosting framework optimized to be highly efficient, effective and better performance [57, 58]. The parameters used for the prediction models are displayed in Table 3. It should be noted that we predicted the presence of defects (binary) in a module and not the number of defects (continuous). The two main experiments conducted are summarized in Figure 1 and described below:

1) Applying NN Filter: we first examined the performance of the defect prediction models when the NN filter procedure (described in Section 3.1) is used in selecting the training data. Considering 14 test and 20 training datasets, we conducted 14 cross-project experiments where the training datasets are merged and used by the test data for filtering to extract the NN modified training data. Following the procedure used by the original creators of the NN filter [11], we used $k = 10$ for the NN filter experiments. The models are trained on the filtered datasets. After model training, the model is tested on a separate test dataset and evaluated using the performance measures discussed in Section 4.3. Our experiment algorithm is shown in Algorithm 1. The NN filter experiment starts from step 1 and stops at step 6, with a sequential flow. The process continues from step 10 through to step 13 where the classifiers are constructed on the filtered $P_{filter}$ instead of the resampled data $P_{syn}$.

2) Resampled Datasets for CPDP: similar to experiment 1, we only modify the training data after applying the NN filter. For the second experiment, the five oversampling approaches discussed in Section 3.2 (i.e. ROS, MAHAKIL, SMOTE, Borderline-SMOTE and

Table 1. Summary of 34 systems extracted from PROMISE data repository

| # | Release | No. of modules | No. of defects | Defects (%) |
|---|---|---|---|---|
| | (Target Data) | | | |
| 1 | Arc | 234 | 27 | 11.5 |
| 2 | Berek | 43 | 16 | 37.2 |
| 3 | e-learning | 64 | 5 | 8 |
| 4 | Intercafe | 27 | 4 | 14.8 |
| 5 | Kalkulator | 27 | 6 | 22.2 |
| 6 | nieruchomosci | 27 | 10 | 37 |
| 7 | Pdftranslator | 33 | 15 | 45.5 |
| 8 | Redaktor | 176 | 27 | 15.3 |
| 9 | Serapion | 45 | 9 | 20 |
| 10 | Skarbonka | 45 | 9 | 20 |
| 11 | Systemdata | 65 | 9 | 13.8 |
| 12 | Tomcat | 858 | 77 | 9 |
| 13 | Workflow | 39 | 20 | 51.3 |
| 14 | Zuzel | 29 | 13 | 44.8 |
| | (Source Data) | | | |
| 15 | ant-1.3 | 125 | 20 | 16 |
| 16 | ant-1.4 | 178 | 40 | 22.5 |
| 17 | ant-1.5 | 293 | 32 | 10.9 |
| 18 | ant-1.6 | 351 | 92 | 26.2 |
| 19 | camel-1.0 | 339 | 13 | 3.8 |
| 20 | camel-1.4 | 872 | 145 | 16.6 |
| 21 | camel-1.6 | 965 | 188 | 19.5 |
| 22 | ivy-1.1 | 111 | 63 | 56.8 |
| 23 | ivy-1.4 | 241 | 16 | 6.6 |
| 24 | ivy-2.0 | 352 | 40 | 11.4 |
| 25 | jedit-4.0 | 306 | 75 | 24.5 |
| 26 | jedit-4.1 | 312 | 79 | 25.3 |
| 27 | jedit-4.2 | 367 | 48 | 13.1 |
| 28 | jedit-4.3 | 492 | 11 | 2.2 |
| 29 | pbeans1 | 26 | 20 | 76.9 |
| 30 | pbeans2 | 51 | 10 | 19.6 |
| 31 | synapse-1.0 | 157 | 16 | 10.2 |
| 32 | xalan-2.4 | 723 | 110 | 15.2 |
| 33 | xerces-1.2 | 440 | 71 | 16.1 |
| 34 | xerces-1.3 | 453 | 69 | 15.2 |

---

[3] http://www.spinelis.gr/sw/ckjm

Table 2. Description of static code metrics [30]

| Metrics | Description | Metrics | Description |
|---|---|---|---|
| WMC | Weighted methods per class | CAM | Cohesion among methods of class |
| DIT | Depth of inheritance tree | IC | Inheritance coupling |
| NOC | Number of children | CBM | Coupling between methods |
| CBO | Coupling between objects | AMC | Average method complexity |
| RFC | Response for a class | Ca | Afferent couplings |
| LCOM | Lack of cohesion of methods | Ce | Efferent couplings |
| LCOM3 | Another form of LCOM | CC | McCabe's cyclomatic complexity |
| NPM | Number of public methods | Max(CC) | Max CC values of methods in class |
| DAM | Data access metric | Avg(CC) | Mean CC values of methods in class |
| MOA | Measure of aggregation | LOC | Lines of code |
| MFA | Measure of functional abstraction | Defects | Number of detected bugs in the class |

Table 3. Classification models and their parameter configurations

| Model | Overview | Parameters |
|---|---|---|
| NB | Naive Bayes | Default |
| NNET | Neural Network | Hidden-layer-sizes={30, 30, 30} |
| KNN | K-Nearest Neighbour | n-neighbours=3 |
| RF | Random Forest | n−estimators=(1000), random−state=(42) |
| SVM | Support Vector Machine | kernel=(linear) |
| XGB | Extreme Gradient Boosting | Default |

ADASYN) and three undersampling approaches (i.e. RUS, OSS, Tomek Links) are applied to the filtered dataset to generate exactly the same number of samples for the minority class and reduce the samples of the majority class, respectively, in the dataset. For both SMOTE and ADASYN, we set $k = 5$ as used by the original authors of these approaches [47, 43]. Each data resampling approach is applied separately, and hence, a total of 84 (14 datasets × 6 prediction models) CPDP models are conducted for each resampling approach. Overall, 672 (84 × 8 resampling approaches) CPDPs are conducted across all resampling approaches. Algorithm 1 displays the procedure followed in conducting experiment 2.

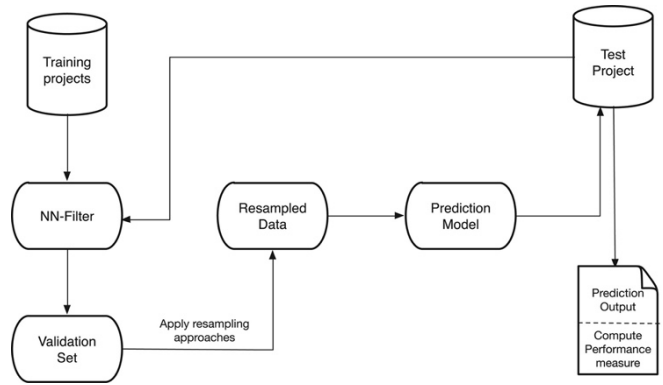

Figure 1. Framework of experiments

*4.4.1 Statistical test and comparison*

As recommended by Kitchenham et al. [59] and Hosseini et al. [16], the application of non-parametric statistical tests and effect sizes to empirical studies in software engineering are beneficial to produce more practical and relevant results for valid conclusions and insights. We compare the results of the performance measures using the original NN-filtered dataset (represented as NOS) and that of using the resampled data (NN-filtered datasets modified with data resampling approaches). To analyse the statistical significance of the prediction performance of models trained on the original NN filter and resampled dataset experiments, Brunner's statistical test [28] is adopted for pairwise comparison. Across all four performance measures, the Win-Tie-Loss values are computed for each pairwise comparison and presented to clearly show any significant difference in prediction performances. Brunner's statistical test was recently recommended as a more robust alternative to the t-test for empirical studies in software engineering [59]. For each selected prediction model and training dataset, Brunner's test is performed

```
Algorithm 1 Outline of the experiments
1  Procedure Begin
2  while Q_test ≠ empty (i.e., testing data is not exhausted) do
3      if source projects P_source{P_1, . . ., P_n} and target projects
          Q_test{Q_1, . . ., Q_j}, j < n then
4          Initialize auc, pf, pd, g-mean ∅; resampling=[MAHAKIL,
           SMOTE, ADASYN, ROS, BORDERLINE-SMOTE, RUS,
           OSS, TOMEK];
5          Initialize filter={NN-filter}, learner=[NB, SVM, KNN, NNET,
           RF, XGB]
6          Merge all training datasets(P_source) into one to obtain P_all
7          for i = 1,..........., length(Q_test) do
8              Filtered P_filter = filter(P_all, Q_test[i])
9              for k = 1,..........., length(sampling) do
10                 Generate resampled dataset (P_syn) from P_filter where
                   P_syn = resampled(P_filter, resampling[k] );
11                 for l = 1,..........., length(learner) do
12                     Train classifier learner_l on resampled data P_syn
13                     Classify data points in Q_test[i] using the classifier
                       learner_l
14                     Compute: auc, pf, pd, g-mean for each classifier
                       learner_l
15                 end
16             end
17         end
18     end
19 end
20 End Procedure
```

across the cross-project pairs for each performance measure and the Win, Tie or Loss value is reported based on the significant value at $p < 0.05$ two-tailed test.

We also apply Cliff's effect size with Hochberg's method proposed by the authors in [28], which measures the size (no effect, small, medium, large) of values between two distributions. The magnitude labels for the effect sizes ($\hat{\delta}$) are interpreted using the thresholds (small [$\hat{\delta} \leq 0.112$], medium [$0.112 > \hat{\delta} < 0.428$] and large $\hat{\delta} \geq 0.428$]) proposed in [60].

## 5. RESULTS

Here, we present our results, which is presented based on the three research questions we address in this paper.

### 5.1 RQ1: What is the impact of data resampling approaches on NN-filtered datasets in CPDP?

For better demonstration, the results are presented in Figure 2 using quartile plots. The quartile plot displays the variation of the results produced by each resampling approach for each prediction model on the 14 datasets. For each performance measure, the quartile plot compares the performance of each resampling approach for each prediction model. These plots are generated by sorting the performance values for all 14 test datasets to isolate the median, lower and upper quartiles. In the performance results in Figure 2, the median is displayed by a solid dot and the quartile limits represent the 25th and 75th percentiles.

Higher medians denote better performance for all performance measures except the $pf$ measure, where lower median values denote better performance. The AUC performance values were comparable for all prediction models and resampling approaches. Data resampling approaches did not significantly improve AUC prediction performance values. Prediction models trained on resampled datasets produced better result values on average for the $g$-$mean$ and recall($pd$) values but performed poorly for the $pf$ performance measure.

Models trained on the resampled datasets attained higher recall ($pd$) and $g$-$mean$ values over models trained on the default NN-filtered datasets (represented as NOS). All oversampling methods and the RUS method improved the pd and $g$-$mean$ values. Tomek links and One-Sided Selection (OSS) approaches did not significantly improve the $pd$ and $g$-$mean$ values compared to the default NN-filtered datasets (NOS) results. Models trained on the default NN-filtered datasets (NOS) performed well with very low $pf$. The models trained on resampled datasets, however, attained high $pf$ comparatively. This confirms, the conclusion previously made by Turhan et al. [11], that higher $pd$ values are accompanied by higher pf values. Most of the oversampling methods and the RUS undersampling method result in higher $pf$ across all models except NB. We observe that no single prediction model performed best for all target datasets considering all the performance measures. This could be explained by the difference in class distribution within the target datasets.

### 5.2 RQ2: What are the high-performing resampling approaches for NN-filtered datasets in CPDP?

To find the statistical significant differences between the performances of the default NN-filtered datasets (NOS) and resampled datasets on the prediction models, total aggregated win-tie-loss values from Brunner's statistical test are presented in Tables 4 and 5. Due to space limitations, we present the top 14 and bottom 14 per each performance measure. Considering the AUC performance measure, we observe from Table 4 that the use of data resampling approaches could not improve the CPDP for all models. Out of the 54 pairwise Win-Tie-Loss comparisons, all resampling approaches combined with the Naive Bayes classifier were found in the bottom half with Borderline-SMOTE and RUS outperforming the other resampling approaches. The majority of the tests resulted in ties and very few losses. Notwithstanding, resampling approaches performed worse for the Random Forest and XGBoost models. However, we observe that the losses are few compared to the three performance measures ($g$-$mean$, $pd$ and $pf$). Considering the g-mean and recall ($pd$) win-tie-loss comparison results in Table 4, the prediction performances on the resampled datasets were statistically significant and different from the prediction performances on the original NN-filtered data. The models trained on the default NN-filtered data were always found in the bottom half of the table indicating that data resampling approaches were always better than no sampling method.

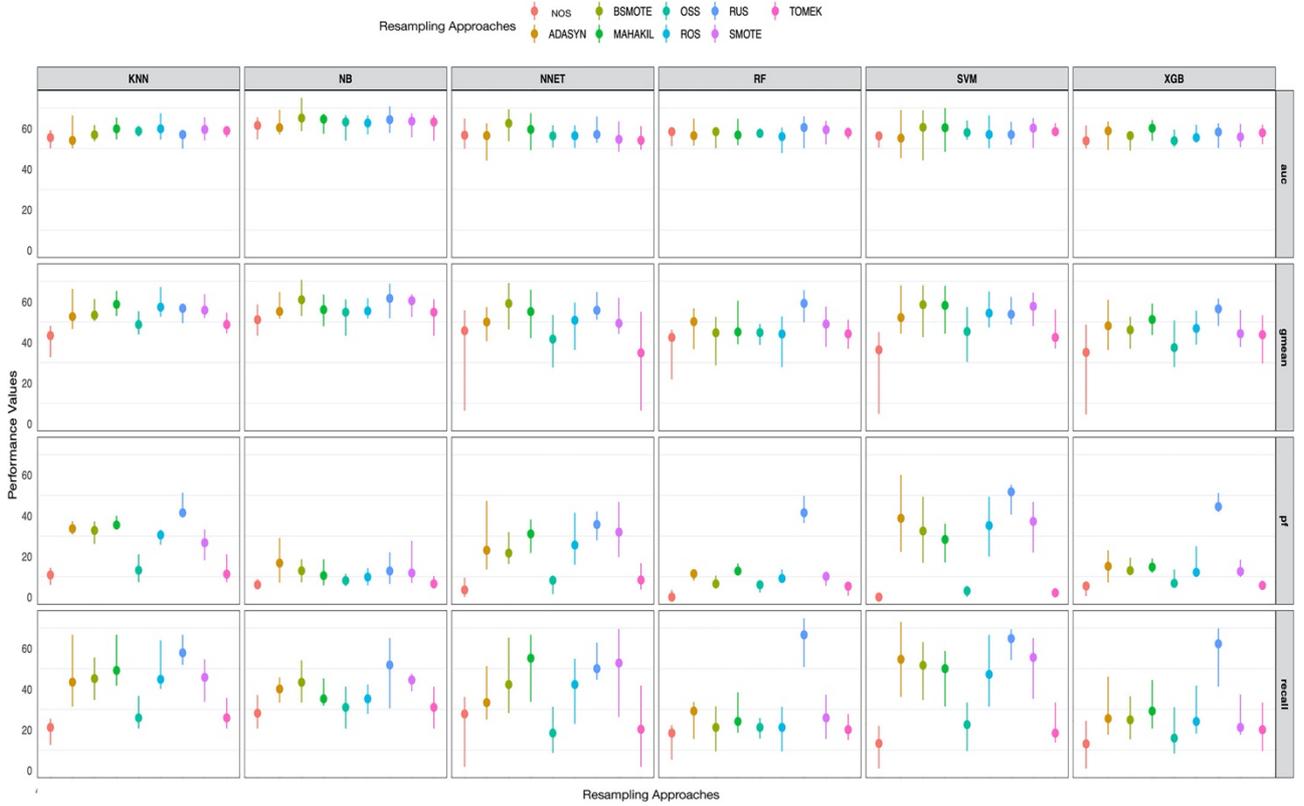

Figure 2. Quartile plots of performance values for all resampling approaches on the filtered datasets per different predictive models (across all 14-test datasets)

However, the models trained on the default NN-filtered datasets (NOS) were significantly improved regarding the $pf$ performance measure, thus outperforming most data resampling approaches. The next best performing data resampling approaches were the undersampling methods Tomek links and OSS. Random under and oversampling, SMOTE and ADA-SYN recorded less than three wins with the highest number of losses.

Furthermore, we compute the overall effect sizes after conducting the win-tie-loss statistical tests across all datasets and present the results in Figure 3. As shown in Figure 3, the performance values of the default models (NOS) is compared to each resampling approach across all datasets and the win- tie-loss statistic is recorded. Additionally, the practical effect (effect sizes) of the statistical results are also computed and presented in Figure 3. For each pairwise comparison, the wins, ties or losses (represented with square shapes) and the magnitude of the effect size (represented with circle shapes) across all datasets per each performance measure are presented.

Considering the AUC performance measure, we observe from Figure 3 that the use of data resampling approaches could not improve the cross-project prediction performance for all models. Out of the pairwise Win-Tie-Loss comparison, the majority of the tests resulted in ties and only two wins for MAHAKIL and one win for BSMOTE. The wins were, however, not practically significant (no effect). There were no losses. This is in agreement with prediction performance in within-project defect prediction where the AUC prediction performance of models trained on resampled datasets was not significantly different from that trained on default datasets [21].

Considering the $g\text{-}mean$ and recall ($pd$) win-tie-loss comparison results in Figure 3, the prediction performances on the resampled datasets were statistically significant and different from the prediction performances on the original NN filtered data. The data resampling approaches specifically the oversampling and Random undersampling approaches significantly outperformed the NOS method mostly achieving large effect sizes (green square and red circle). For most pairwise comparisons, the data resampling approaches achieved significant (win) statistical tests and few no effects as the effect size status. However, the models trained on the default NN-filtered datasets (NOS) were significantly better regarding the pf performance measure, thus outperforming most data resampling approaches. In agreement with the results in Table 4, the next best performing data resampling approaches were the undersampling approaches Tomek links and OSS. This indicates that the combining data resampling approaches with the original NN-filtered data should be avoided if we aim at achieving lower pf values while the data resampling approaches should be considered when higher $pd$ and $g\text{-}mean$ values are required.

### 5.3 RQ3: Is the application of data resampling approaches practical and beneficial for CPDP?

To assess the practical benefits of data sampling approaches, the actual number of test sets that meet the practical criteria of 75% for all performance measures except pf, where the criterion is 15%, is summarised in Figure 4. For each performance measure, after 756 (14 × 6 × 9) runs across all 14 test sets, six prediction models and nine sampling methods, the success rates were very low. All data resampling approaches attained success rates of less than 10% for the AUC and g-mean performance

Table 4. Performance in terms of wins, losses, and wins-losses aggregated from the predictors (model resampling method) on the 14 datasets per each performance measure

| | AUC | | | | | | g-mean | | | | | |
|---|---|---|---|---|---|---|---|---|---|---|---|---|
| | Model | Resampling approach | Wins | Losses | Wins–Losses | Ties | Model | Resampling approach | Wins | Losses | Wins–Losses | Ties |
| 1 | NB | BORDERLINE | 36 | 0 | 36 | 17 | NB | BORDERLINE | 32 | 0 | 32 | 21 |
| 2 | NB | RUS | 32 | 0 | 32 | 21 | NB | RUS | 30 | 0 | 30 | 23 |
| 3 | NB | ROS | 20 | 0 | 20 | 33 | NNET | BORDERLINE | 27 | 0 | 27 | 26 |
| 4 | NB | SMOTE | 20 | 1 | 19 | 32 | RF | RUS | 27 | 0 | 27 | 26 |
| 5 | NB | OSS | 17 | 0 | 17 | 36 | KNN | MAHAKIL | 26 | 0 | 26 | 27 |
| 6 | NB | MAHAKIL | 16 | 0 | 16 | 37 | NB | SMOTE | 26 | 0 | 26 | 27 |
| 7 | NB | TOMEK | 13 | 0 | 13 | 40 | SVM | SMOTE | 25 | 0 | 25 | 28 |
| 8 | NNET | BORDERLINE | 11 | 0 | 11 | 42 | KNN | ROS | 24 | 0 | 24 | 29 |
| 9 | NB | NOS | 9 | 1 | 8 | 43 | KNN | SMOTE | 24 | 0 | 24 | 29 |
| 10 | XGB | MAHAKIL | 6 | 1 | 5 | 46 | NB | ADASYN | 25 | 1 | 24 | 27 |
| 11 | NB | ADASYN | 5 | 1 | 4 | 47 | SVM | MAHAKIL | 24 | 0 | 24 | 29 |
| 12 | KNN | MAHAKIL | 3 | 0 | 3 | 50 | KNN | BORDERLINE | 23 | 1 | 22 | 29 |
| 13 | KNN | ROS | 2 | 0 | 2 | 51 | NB | MAHAKIL | 22 | 0 | 22 | 31 |
| 14 | SVM | TOMEK | 4 | 2 | 2 | 47 | SVM | ROS | 21 | 0 | 21 | 32 |
| . | . | . | . | . | . | . | . | . | . | . | . | . |
| 41 | RF | TOMEK | 0 | 6 | −6 | 47 | NNET | NOS | 0 | 23 | −23 | 30 |
| 42 | KNN | RUS | 0 | 7 | −7 | 46 | RF | BORDERLINE | 2 | 25 | −23 | 26 |
| 43 | RF | BORDERLINE | 0 | 7 | −7 | 46 | XGB | BORDERLINE | 2 | 25 | −23 | 26 |
| 44 | XGB | TOMEK | 0 | 7 | −7 | 46 | XGB | TOMEK | 2 | 25 | −23 | 26 |
| 45 | NNET | ADASYN | 0 | 8 | −8 | 45 | NNET | OSS | 0 | 27 | −27 | 26 |
| 46 | RF | OSS | 0 | 8 | −8 | 45 | RF | OSS | 2 | 29 | −27 | 22 |
| 47 | XGB | ROS | 0 | 9 | −9 | 44 | RF | TOMEK | 2 | 30 | −28 | 21 |
| 48 | KNN | NOS | 0 | 10 | −10 | 43 | NNET | TOMEK | 0 | 29 | −29 | 24 |
| 49 | XGB | BORDERLINE | 0 | 10 | −10 | 43 | KNN | NOS | 0 | 30 | −30 | 23 |
| 50 | XGB | RUS | 0 | 10 | −10 | 43 | XGB | OSS | 0 | 33 | −33 | 20 |
| 51 | RF | NOS | 0 | 11 | −11 | 42 | RF | ROS | 1 | 35 | −34 | 17 |
| 52 | XGB | OSS | 0 | 13 | −13 | 40 | SVM | NOS | 0 | 35 | −35 | 18 |
| 53 | RF | ROS | 0 | 14 | −14 | 39 | RF | NOS | 0 | 45 | −45 | 8 |
| 54 | XGB | NOS | 0 | 14 | −14 | 39 | XGB | NOS | 0 | 45 | −45 | 8 |

*Note*: Higher wins (%), wins-losses (%) and lower losses (%) indicate the higher performance of the predictor. The predictors are ordered by wins-losses.

measures. Similarly, all the resampling approaches excluding the RUS and ADASYN also attained success rates of less than 10% for the recall (pd) performance measure. RUS attained the highest success rate (22.6%) considering the pd performance measure, whereas OSS attained the lowest success rate (1.19). The default NN filter approach (NOS) attained lower success rates for pd, AUC and $g\text{-}mean$ than the oversampling methods indicating that data resampling methods with the NN filter result in better performance. However, for false alarms (pf), the default NN filter attained the highest success rate closely followed by OSS and TOMEK links. RUS performed worse for the pf measure with the oversampling methods attaining higher success rates than RUS.

In summary, our results show the following: (1) The use of data resampling approaches significantly improved prediction performance ($pd$, $g\text{-}mean$). Considering the number of ties and losses, there was a statistical difference between the performances of the prediction models trained on resampled data and the original filtered data. (2) To attain better pf values, the default NN-filtered data should not be combined with data resampling approaches. (3) Similar to observations from WPDP [21, 39], the performance of AUC is not significantly impacted by data resampling methods for CPDP models.

## 6. DISCUSSIONS

The experimental results indicate that data resampling approaches significantly impact the performance of CPDP models. Comparatively, the oversampling approaches outperformed the undersampling approaches regarding the $g\text{-}mean$ and $pd$ performance values. The experimental results show that the NN filter alone (NOS) obtained very low practical results, suggesting data sampling methods are needed in addition to the NN filter method to obtain more

Table 5. Performance in terms of wins, losses, and wins-losses aggregated from the predictors (model resampling approach) on the 14 datasets per each performance measure

| | | *pd* | | | | | | *pf* | | | | |
|---|---|---|---|---|---|---|---|---|---|---|---|---|
| | Model | Resampling approach | Wins | Losses | Wins–Losses | Ties | Model | Resampling approach | Wins | Losses | Wins–Losses | Ties |
| 1 | RF | RUS | 43 | 0 | 43 | 10 | SVM | NOS | 52 | 0 | 52 | 1 |
| 2 | SVM | RUS | 41 | 0 | 41 | 12 | RF | NOS | 50 | 0 | 50 | 3 |
| 3 | KNN | RUS | 38 | 0 | 38 | 15 | SVM | TOMEK | 49 | 1 | 48 | 3 |
| 4 | XGB | RUS | 36 | 0 | 36 | 17 | SVM | OSS | 44 | 1 | 43 | 8 |
| 5 | KNN | MAHAKIL | 35 | 1 | 34 | 17 | RF | TOMEK | 42 | 2 | 40 | 9 |
| 6 | SVM | ADASYN | 34 | 0 | 34 | 19 | NB | NOS | 39 | 3 | 36 | 11 |
| 7 | NNET | RUS | 32 | 1 | 31 | 20 | XGB | NOS | 39 | 3 | 36 | 11 |
| 8 | NNET | MAHAKIL | 30 | 1 | 29 | 22 | RF | OSS | 38 | 3 | 35 | 12 |
| 9 | SVM | SMOTE | 29 | 0 | 29 | 24 | XGB | TOMEK | 34 | 4 | 30 | 15 |
| 10 | NNET | BORDERLINE | 28 | 0 | 28 | 25 | RF | BORDERLINE | 35 | 6 | 29 | 12 |
| 11 | NNET | SMOTE | 28 | 0 | 28 | 25 | NB | TOMEK | 32 | 4 | 28 | 17 |
| 12 | SVM | BORDERLINE | 29 | 1 | 28 | 23 | NNET | OSS | 30 | 4 | 26 | 19 |
| 13 | KNN | ADASYN | 28 | 1 | 27 | 24 | NB | OSS | 31 | 6 | 25 | 16 |
| 14 | KNN | ROS | 29 | 3 | 26 | 21 | NNET | NOS | 29 | 6 | 23 | 18 |
| . | . | . | . | . | . | . | . | . | . | . | . | . |
| 41 | SVM | OSS | 2 | 26 | −24 | 25 | NNET | MAHAKIL | 4 | 34 | −30 | 15 |
| 42 | NNET | NOS | 0 | 25 | −25 | 28 | NNET | SMOTE | 3 | 33 | −30 | 17 |
| 43 | KNN | NOS | 3 | 30 | −27 | 20 | KNN | BORDERLINE | 4 | 35 | −31 | 14 |
| 44 | NNET | OSS | 1 | 29 | −28 | 23 | SVM | BORDERLINE | 3 | 34 | −31 | 16 |
| 45 | SVM | TOMEK | 2 | 31 | −29 | 20 | KNN | ADASYN | 4 | 37 | −33 | 12 |
| 46 | XGB | TOMEK | 3 | 32 | −29 | 18 | SVM | SMOTE | 2 | 35 | −33 | 16 |
| 47 | RF | BORDERLINE | 2 | 32 | −30 | 19 | KNN | MAHAKIL | 4 | 38 | −34 | 11 |
| 48 | RF | OSS | 2 | 33 | −31 | 18 | NNET | RUS | 1 | 35 | −34 | 17 |
| 49 | RF | TOMEK | 1 | 34 | −33 | 18 | SVM | ROS | 1 | 35 | −34 | 17 |
| 50 | RF | ROS | 1 | 36 | −35 | 16 | SVM | ADASYN | 0 | 37 | −37 | 16 |
| 51 | XGB | OSS | 0 | 36 | −36 | 17 | RF | RUS | 0 | 44 | −44 | 9 |
| 52 | SVM | NOS | 0 | 42 | −42 | 11 | KNN | RUS | 0 | 45 | −45 | 8 |
| 53 | RF | NOS | 0 | 46 | −46 | 7 | SVM | RUS | 0 | 47 | −47 | 6 |
| 54 | XGB | NOS | 0 | 46 | −46 | 7 | XGB | RUS | 0 | 48 | −48 | 5 |

*Note*: Higher wins (%), wins-losses (%) and lower losses (%) indicate the higher performance of the predictor. The predictors are ordered by wins-losses.

significant and practical results. Conventional sampling approaches such as the SMOTE, RUS and ROS produced significant and improved performance results but were accompanied with a high false alarm rate.

Predictors that generalize well on training data are known to perform better than models trained on restricted/few data samples. Training prediction models on very large (infinite) training datasets that capture all variations in the data distribution is ideal for higher and improved prediction performance. Due to the inability to obtain infinite training datasets, it has been recommended to extend the training set with artificially created examples that increases the diversity within the data distribution [61]. It was also suggested that a variety of (diverse) data samples will significantly improve the prediction performance. However, this was not the case for CPDP as the diversity-based oversampling approach (MAHAKIL), although improving performance, could not outperform the more conventional approaches, which do not necessarily increase the diversity within the dataset. This is because MAHAKIL aims to carefully generate diverse minority class instances in a specified region reducing the false alarms. However, with the NN Filter and CPDP whereby data samples are selected from different projects, the final obtained datasets for training will be by default diverse. MAHAKIL will thus generate very few or limited diverse instances and this affects the performance. Nevertheless, MAHAKIL outperforms the other oversampling approaches regarding lower $pf$ values. Similarly, TOMEK and OSS undersampling approaches significantly resulted in lower $pf$ values compared to the traditional RUS approach. Quality assurance teams will prefer MAHAKIL, TOMEK and OSS over the other conventional sampling approaches since low pf values are preferred by these teams.

Additionally, our analysis revealed that, although the performance of data resampling approaches on the CPDP

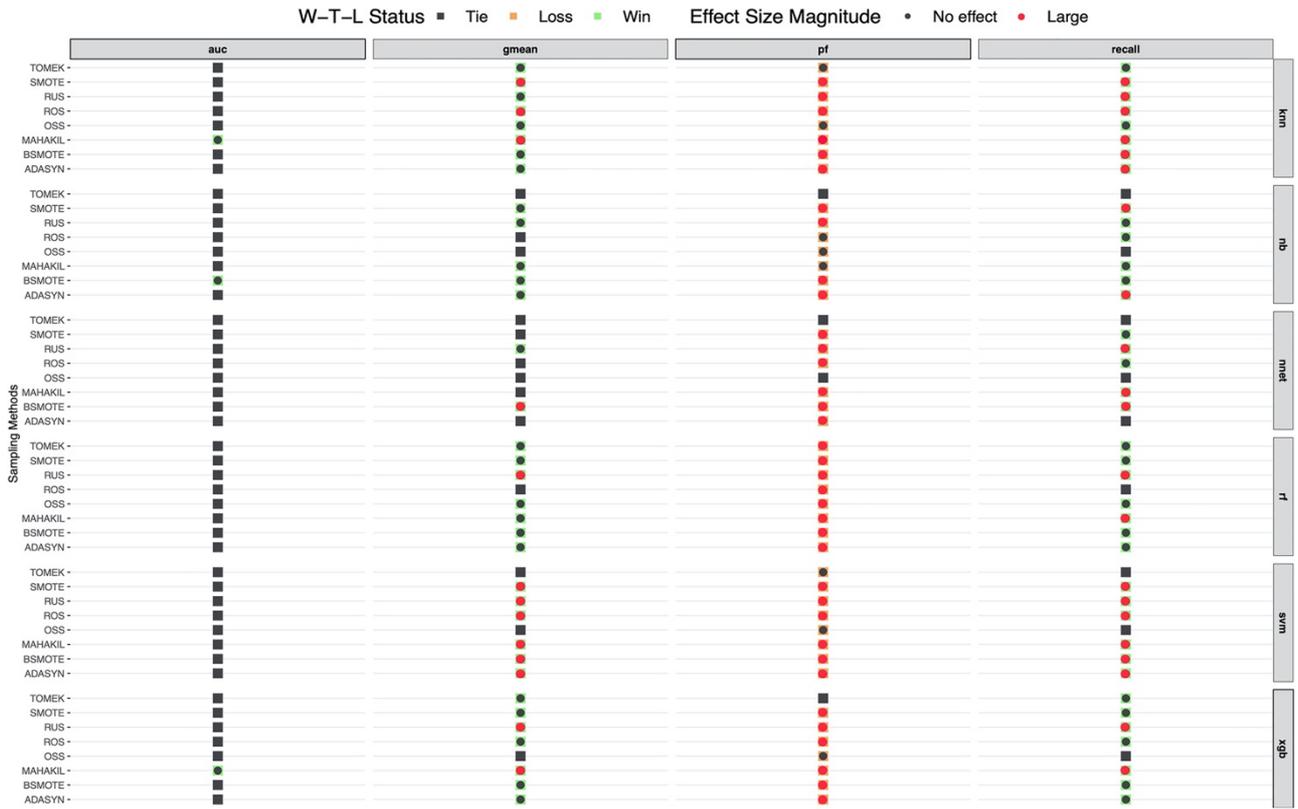

Figure 3. Brunner's statistical test win-tie-loss and effect-size comparison of NOS versus ROS, ADASYN, Borderline-SMOTE, SMOTE, MAHAKIL, OSS and TOMEK across all datasets per each defect prediction model and performance measure (AUC, $g\text{-}mean, pd, pf$)

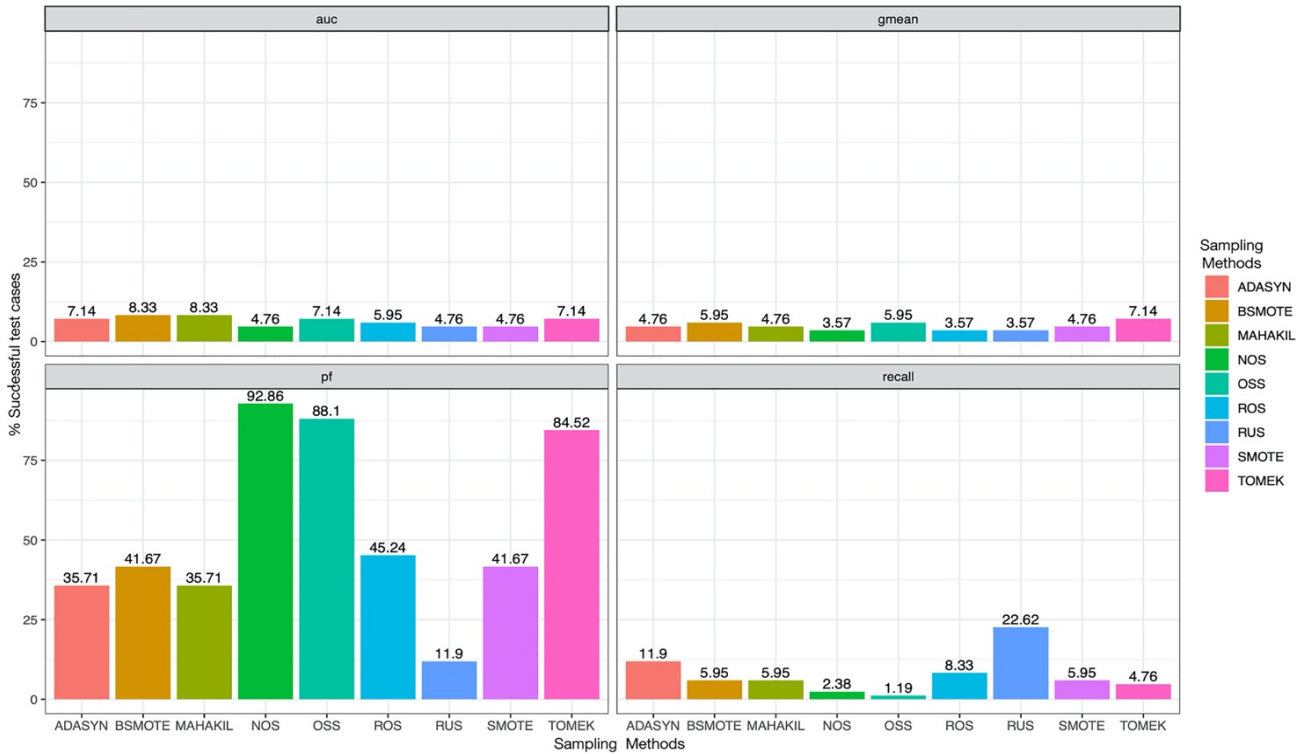

Figure 4. Measuring the practical success of sampling approaches on cross-project defect prediction by the percent of test sets whose defects were predicted and met the criteria for pd, AUC and g-mean greater than 75% and pf less than 15%

models did improve the AUC values, the values were not too high, ranging between 45% and 65% across all models (see Figure 2). This is confirmed by the low success rates (<10%) obtained as presented in Figure 4. Data resampling approaches, especially the oversampling approaches, with the Naive Bayes model produced the highest AUC values.

The AUC performance metric is useful in ranking or prioritizing the crucial target samples [62], which is the defective modules in our experiment. A perfect AUC value of 100% implies that all modules are efficiently ranked—assigning higher priorities to the defective modules than the non-defective modules. An average AUC value of 50% indicates that a random prioritization is no worse than the prioritization using the advanced model (resampled dataset and prediction model). With very small practical AUC values, the use of data resampling approaches for prioritization of modules for quality assurance activities (such as testing and code review) is not recommended for software projects. With large $pf$ values, the selection of top $k$% faulty modules will always be contaminated with the non-defective modules. Notwithstanding, data resampling approaches have a silver lining. They can aid in classifying all defective modules.

## 7. THREATS TO VALIDITY

As an empirical study, there are several potential limitations. We discuss below the internal and external threats to the validity of the study. The main external limitation of this study is that it is dependent on the NN filter. All experiments were conducted using the NN filter as the preprocessor of the dataset. As stated in the introduction, the NN filter has been widely evaluated as having a positive impact on CPDP models. Studies without the use of the NN filter is left as future work. Considering only open-source software projects with source code metrics poses an external threat. These metrics extracted with automated software are easy to collect but the results cannot be generalized to all other projects that have different metrics apart from static code or to projects in the commercial domain. Also, the number of projects considered could have an effect on our results. We, however, considered a wide range and sufficient sizes of projects and, therefore, produced robust results, which we expect to be similar for other unconsidered projects. We intend to consider commercial projects in future studies. The selection of our source and target data also poses a threat to our results. Since we only considered single-version projects as our target projects, it is unclear if our results would generalize to the other datasets, which were not used as the source data. Further work where every project is considered as a target project is required to clarify this.

A limited number of predictive models were also considered in this paper. Considering different types of predictive models could have different implications. However, these are widely used models in past studies. Similarly, the number of neighbours chosen for the NN filter had an impact on our results. Exploring large values for these neighbours could lead to different results since the filtered dataset sizes would be affected. We considered four major performance measures, which are widely used for defect prediction on class imbalanced datasets. Other evaluation measures such as the effort-aware measures could be considered in future studies.

## 8. CONCLUSION

This paper investigates the impact of using data sampling methods for improving the performance of CPDP models. Employing eight data resampling approaches, we resampled the datasets and used them in training CPDP models after filtering the training data using the NN filter. The data resampling methods did improve the prediction performance as expected by significantly improving the recall ($pd$) and $g\text{-}mean$ performance measures. The use of the default NN-filtered dataset was significantly better than the resampled data in terms of the false alarms ($pf$). However, data resampling methods did not improve the AUC performance values. For practical results, AUC and $g\text{-}mean$ results of all models attained success rates of less than 10%. Random undersampling improved the recall values but that was accompanied with high false alarms. The experiments demonstrated that data resampling methods could mitigate the negative effects of class imbalance on the performance of CPDP in terms of improving the probability of detection (recall).

From these results, we have demonstrated that data resampling methods should be adopted for constructing prediction models in a bid of reducing the negative effects of class imbalance within a dataset obtained from several projects. This study lays the foundation for further future works. We aim to investigate the use of cost-sensitive techniques for CPDP. With a low practical success rate (less than 10%), a further empirical study is required to investigate how prediction performance can be improved using other methods.

**ACKNOWLEDGEMENT**
This work has in parts been supported by ELLIIT; the Swedish Strategic Research Area in IT and Mobile Communications.

**CONFLICT OF INTEREST**
None.

**DATA AVAILABILITY STATEMENT**
The data that support the findings of this study are openly available in PROMISE Repository at https://zenodo.org/communities/seacraft.